\documentclass[pra,twocolumn]{revtex4-2}
\usepackage[T1]{fontenc}
\usepackage{array}
\usepackage{float}
\usepackage{multirow}
\usepackage{amsmath}
\usepackage{amssymb}
\usepackage{graphicx}
\usepackage{epsfig,amsmath}
\usepackage{booktabs}
\usepackage{subfigure}
\usepackage{dcolumn}
\usepackage{stmaryrd}
\usepackage{mathrsfs}
\usepackage{pifont}
\usepackage{bm}
\usepackage{latexsym}
\usepackage{hyperref}
\usepackage{color}
\usepackage{newfloat,algcompatible}  
\usepackage[size=small]{caption}
\usepackage{etoolbox}
\usepackage{newfloat}
\pdfpageheight\paperheight
\pdfpagewidth\paperwidth

\floatstyle{ruled}
\newfloat{algorithm}{H}{loa}
\providecommand{\algorithmname}{Algorithm}
\floatname{algorithm}{\protect\algorithmname}
\makeatother
\begin{document}
\title{Deep reinforcement learning for universal quantum state preparation via dynamic pulse control}
\author{Run-Hong He$^{1}$, Rui Wang$^{1}$, Jing Wu$^{1}$, Shen-Shuang Nie$^{1}$,  Jia-Hui Zhang$^{1}$ and Zhao-Ming Wang$^{1}$}
\email{mingmoon78@126.com}
\address {$^{1}$College of Physics and Optoelectronic Engineering, Ocean University of China, Qingdao 266100, China}
\date{\today}

\begin{abstract} 
	Accurate and efficient preparation of quantum state is a core issue in building a quantum computer. 
	In this paper, we investigate how to prepare a certain single- or two-qubit target state from arbitrary initial states in semiconductor double quantum dots with the aid of deep reinforcement learning. Our method is based on the training of the network over numerous preparing tasks. The results show that once the network is well trained, it works for any initial states in the continuous Hilbert space. Thus repeated training for new preparation tasks is avoided. Our scheme outperforms the traditional optimization approaches based on gradient with both the higher designing efficiency and the preparation quality in discrete control space.
	Moreover, we find that the control trajectories designed by our scheme are robust against static and dynamic fluctuations, such as charge and nuclear noises.
	
\end{abstract}

\maketitle
\section{Introduction}
\label{Sec.1}
Future quantum computers promise exponential speed-ups over their classical counterparts in solving certain problems like search and simulation \cite{nielsen2002quantum}. A wide variety of promising modalities emerges in the race to realize the quantum computer, such as trapped ions \cite{richerme2014non,casanova2014transistor}, photonic system \cite{bellec2012faithful,perez2013perfect,peruzzo2014variational,chapman2016experimental}, nitrogen-vacancy centers \cite{childress2013diamond}, nuclear magnetic resonance \cite{vandersypen2005nmr}, superconducting circuits \cite{devoret2013superconducting,wendin2017quantum} and semiconductor quantum dots \cite{zajac2018resonantly,huang2019fidelity,watson2018programmable,jang2020three,hanson2007spins,eriksson2004spin,zwanenburg2013silicon}.
Among these the semiconductor quantum dots is a powerful competitor for potential scalability, integrability with existing classical electronics and well-established fabrication technology. Spins of electrons, which are trapped in quantum dots structure based on Coulomb effect, can serve as spin-qubits for quantum information \cite{loss1998quantum}.
Spin qubits can be encoded in many ways, such as spin-1/2, singlet-triplet ($S$-$T_0$) and hybrid systems \cite{zhang2019semiconductor}. 
In particular, the spin $S$-$T_0$ qubit in double quantum dots (DQDs) attracts much attention for the merit that it can be manipulated solely with electrical pulses \cite{taylor2005fault,wu2014two,nichol2017high}.

It has been proved that several arbitrary single-qubit gates plus an entangling two-qubit gate are the prototypes of all other logic gates in quantum algorithm which is implemented on a circuit-model quantum computer \cite{nielsen2002quantum,quantum_computation_2020}. 
In an authentic sense, the implementation of any single- and two-qubit gates can be reduced to the state preparation problems.
Generally, arbitrary manipulations of a single-qubit can be achieved by successive rotations around the $x$- and $z$-axes on the Bloch sphere.
In the context of  $S$-$T_0$ single-qubit in semiconductor QDQs, the only tunable parameter $J$ is the rotation rate around the $z$-axis, while the rotation rate $h$ around the $x$-axis is difficult to be changed  \cite{throckmorton2017fast}. For two coupled $S$-$T_0$ qubits based on electrostatic interaction \cite{taylor2005fault,shulman2012demonstration,nichol2017high,van2011charge}, operations can be achieved by tuning the strength of the  $J_i$ on each qubit, where $i=1,2$ refers to the corresponding qubit. To guarantee the entanglement between two qubits, $J_i>0$ must be kept during the runtime \cite{petta2005coherent,shulman2012demonstration,wang2015improving,nichol2017high}.

Various of schemes have been proposed to add proper pulses on $J$ to control the qubits \cite{wang2012composite,kestner2013noise,wang2015improving}.
It is typically required to iteratively solve a set of nonlinear equations \cite{wang2014robust,yang2018neural} for analytically tailoring the control trajectory, so it is a computationally exorbitant and time-consuming task in practice. There are also several traditional optimal methods based on gradient that can be used to design the control trajectory, such as stochastic gradient descent (SGD) \cite{SGD}, chopped random-basis optimization (CRAB) \cite{CRAB_1,CRAB_2} and gradient ascent pulse engineering (GRAPE) \cite{GRAPE_1,GRAPE_2}. 
However, due to the sensitivity to the initial control trajectory setting, these traditional methods can only find the local optimum \cite{GRAPE_2}. 
In addition, their efficiency is limited by their iterative nature, especially when there exists a large number of preparing tasks. 
Except for these traditional routes, recently the deep reinforcement learning (RL) \cite{goodfellow2016deep} shows a wide applicability in quantum control problems \cite{zhang2018automatic,yang2020optimizing,lin2020quantum,bukov2018reinforcement,bukov2018reinforcement2,kong2020artificial,palmieri2020experimental,RL_cartpole,RL_CNOT,RL_Universal,RL_Grover}. With deep RL, the appropriate control trajectory can be found for the driving between two given states.
For example, how to drive a qubit from a fixed initial state to another fixed target state \cite{RL_CNOT,zhang2019does} with the aid of deep RL \cite{sutton2018reinforcement} has been investigated.

Recently, the generation of arbitrary states from a specific state \cite{ppo_state_preparation} in nitrogen-vacancy center has been realized by leveraging the deep RL. Then it is intriguing to check if the deep RL can be used to realize a contrary problem: preparing a certain target state from arbitrary initial states, i.e., universal state preparation (USP). In practical quantum computation, it is often required to reset an arbitrary state to a specific target state \cite{wang_almost_exact_state_transfer,wang_almost_exact_state_transfer_environment,ren_grover_environment}. 
For example, the initial state of the system always needs to be set at the ground state when transferring a quantum state through a spin chain \cite{wang_almost_exact_state_transfer,wang_almost_exact_state_transfer_environment}. In the realization of quantum Toffoli or Fredkin gate, the ancilla state must be preprepared to the standard state $|0\rangle $ or $|1\rangle $ in certain issues \cite{divincenzo1995two,feynman1982simulating,smolin1996five}. 
Generation of two-qubit entangled state is also required \cite{nielsen2002quantum} in completing quantum information processing tasks, such as the teleportation \cite{bennett1993teleporting,bouwmeester1997experimental}. Note that the network requires being trained again once the preparing task changes \cite{RL_CNOT,zhang2019does}. This could be an exhausting work when there are lots of different states waiting to be prepared to a certain target state. In this paper, we investigate the USP with the aid of deep RL in such a constrained driving parameters system. Benefited from a more sufficient learning on numerous preparing tasks, we find that USP can be achieved with a single training of the network. Evaluation results show that our scheme outperforms the alternative optimization approaches both in terms of the preparation quality and designing efficiency in discrete control space. 
Additionally, the control trajectories exhibit strong robustness against various imperfections that come from the system, environment or control field. We point out that by combining our scheme with Ref.~\cite{ppo_state_preparation}, the preparation of arbitrary states from arbitrary states can be realized.

\section{Models and Methods}
\label{Sec.2}
At first, we present the models of electrically controlled  $S$-$T_0$ single- and two-qubit in semiconductor DQDs in Subsections~\ref{single-qubit in DQDs} and \ref{two-qubit in DQDs}, respectively. 
Then we present our USP algorithm in Subsection~\ref{USP with deep RL}.
\subsection{Voltage-controlled single-qubit in semiconductor DQDs}
\label{single-qubit in DQDs}
The effective control Hamiltonian of a single-qubit encoded by $S$-$T_0$ states in semiconductor DQDs can be written as \cite{malinowski2017notch,petta2005coherent,bluhm2009universal,maune2012coherent},

\begin{equation}\label{eq1}
	H(t)=J(t)\sigma_{z}+h\sigma_{x}.
\end{equation}
It is written under the computational basis states: the spin singlet state $|0\rangle=|S\rangle=(|\uparrow\downarrow\rangle-|\downarrow\uparrow\rangle)/\sqrt{2}$
and the spin triplet state $|1\rangle=|T_0\rangle=(|\uparrow\downarrow\rangle+|\downarrow\uparrow\rangle)/\sqrt{2}$.
Here the arrows indicate the spin projections of the electron in
the left and right dots, respectively. $\sigma_{z}$ and $\sigma_{x}$ are the Pauli matrices. $h$ represents the Zeeman energy spacing of two spins. Considering $h$ is difficult to be changed experimentally \cite{zhang2019semiconductor}, here we assume
it is a constant and set $h=1$. We also take the reduced Planck constant $\hbar=1$ throughout. Physically the exchange
coupling $J(t)$ is tunable, non-negative and finite \cite{zhang2019semiconductor}.

Arbitrary qubit states can be written as
\begin{equation}\label{eq2}
	|s\rangle=\mathrm{cos}\frac{\theta}{2}|0\rangle+e^{i\varphi}\mathrm{sin}\frac{\theta}{2}|1\rangle,
\end{equation}
where $\theta$ and $\varphi$ are real numbers that define points
on the Bloch sphere. For a certain initial state $|s_{ini}\rangle$
on the Bloch sphere, an arbitrary target state $|s_{tar}\rangle$
can be achieved by successive rotations around the $x$- and $z$-axes
of the Bloch sphere. In the context of semiconductor DQDs, $h$ and $J(t)$ cause a rotation around the $x$-axis and $z$-axis of
the Bloch sphere, respectively.
\subsection{Capacitively coupled $S$-$T_0$ qubits in semiconductor double quantum dots}
\label{two-qubit in DQDs}
Operations on two entangled qubits are often required in quantum information processing. In semiconductor DQDs, interqubit operations can be performed on two adjacent and capacitively coupled $S$-$T_0$ qubits. In the  basis of $\{|SS\rangle,|ST_0\rangle,|T_0S\rangle, |T_0T_0\rangle\}$, the Hamiltonian can be written as \cite{shulman2012demonstration,nichol2017high},

\begin{equation}\label{eq3}
	H_{2\!-\!qubit}\!=\!\frac{\hbar}{2}
	\left(\begin{array}{cccc}
		J_{1}\!+\!J_{2} & h_{2}           & h_{1}           & 0                          \\
		h_{2}           & J_{1}\!-\!J_{2} & 0               & h_{1}                      \\
		h_{1}           & 0               & J_{2}\!-\!J_{1} & h_{2}                      \\
		0               & h_{1}           & h_{2}           & -J_{1}\!-\!J_{2}\!+\!2J_{12}
	\end{array}\right),
\end{equation}
where $h_i$ and $J_i$ are the Zeeman energy spacing and exchange coupling of the $i$th qubit  respectively. $J_{12}\propto J_1J_2 $ refers to the strength of Coulomb coupling between two qubits. $J_i>0$ is required to maintain the interqubit coupling all the time. For simplicity, we take $h_1=h_2=1$ and $J_{12}=J_1J_2/2$ here.

\subsection{Universal state preparation via deep reinforcement learning} 
\label{USP with deep RL}

Our target is to drive arbitrary initial states to a certain target state via dynamical control pulses. 
The control trajectory is discretized as a piece-wise constant function \cite{GRAPE_1}. This control field can be experimetally realized with an arbitrary wave generator \cite{superconducting_guide,nichol2017high,petta2005coherent,shulman2012demonstration,wang2015improving}. The strategy used here is to generate this control trajectory with the deep Q network algorithm (DQN) \cite{mnih2013playing,mnih2015human}, which is an important member of deep RL. The details of the DQN are described in Section~\ref{Sec.5}. Here we just refer it as a neural network, i.e., the Main Net $\theta$.  

\begin{figure}[ht] 
	   \centering
		\includegraphics[width=8cm]{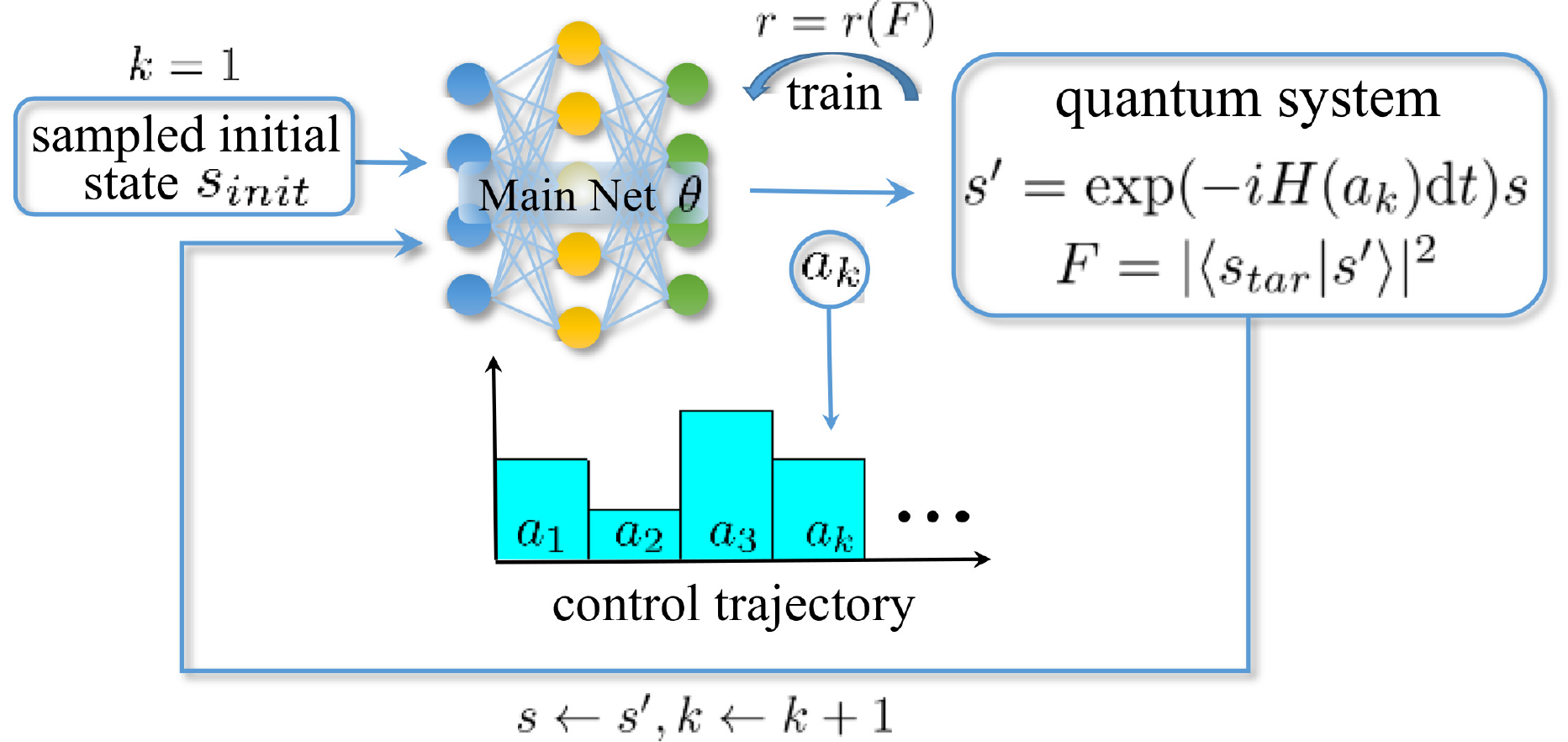}
	\caption{Overview of the USP algorithm to learn and to practice the control trajectory designing. The details of the algorithm is described in the main text of Subsection~\ref{USP with deep RL}  and in the pseudocode Algorithm~\ref{algo}.}
	\label{Fig1}
\end{figure}

Firstly, a database comprised of numerous potential initial quantum states is divided into the training set and the test set. Secondly, all the states in the training set will be used to train the Main Net in turn.  The random-initialized Main Net is initially fed with a sampled initial state $s$ at time step $k=1$. Then outputs the predicted ``best action'' $a_k$. According to the current state $s$ and the action $a_k$, calculate the next state $|s'\rangle=\mathrm{exp}(-iH(a_k)\mathrm{d}t)|s\rangle$ and the corresponding fidelity $F=|\langle s_{tar}|s'\rangle|^2$. The fidelity $F$ indicates how close the next state is to the target state. Then the next state $s'$ is fed to the Main Net as the current state with time step $k\leftarrow k+1$. The fidelity will be enveloped as the reward $r=r(F)$ and be used to train the Main Net. Now repeat the above operations until the episode terminates when $k$ reaches the maximum time step or the fidelity excesses a certain satisfactory threshold. Correspondingly, the control trajectory is constructed by these predicted actions orderly. This single training scheme has been used to realize the driving between two individual states \cite{zhang2019does}. For our training scheme, we continue to train the Main Net in the same way as above with the other initial states in the training set in turn. By sufficient training of more preparing tasks, the Main Net learns to assign an action-value (also named $Q$-value) to each state and action pair gradually according to the correspondence between them and the target state. With accurate $Q$-values, it is easily to determinate which action should be chosen in a given state.  So that the Main Net can match every potential state with a reasonable action towards the target state. Finally, the well-trained Main Net can be used to tailor the appropriate control trajectories for these initial states databased in the test set and even all other states in the continuous Hilbert space. The overview of this training and designing process is pictured in Figure~\ref{Fig1}.  
And a full description of the training process is given in Algorithm~\ref{algo}.

\section{Results}
\label{Sec.3}
In this section, we compare and contrast the performance of our scheme with two sophisticated optimization approaches based on gradient in preparing a certain target state from arbitrary initial states. We exemplarily consider the preparation of two standard single-qubit states: $|0\rangle $ and $|1\rangle$ as well as a two-qubit state: the Bell state $(|00\rangle+|11\rangle)/\sqrt{2}$.

\subsection{Universal single-qubit state preparation}
\label{single-qubit state preparation}

Now we consider the preparation of two single-qubit states $|0\rangle$ and $|1\rangle$ by using our USP algorithm. Here the discrete allowed actions $J(t)$ are empirically set to be 0, 1, 2 or 3 with duration $\mathrm{d}t=\pi/40$. We stress that these settings can also be further tailored as required.
The Main Net consists of an input layer, an output layer and
two hidden layers with 4, 4, 20 and 20 neurons respectively. 
The maximum total operation time is limited to be $\pi$, which is uniformly discretized into 40 slices.
The reward function should be set to allow a rapid growth in itself as the fidelity increases and a big bonus is given when the fidelity reaches a desired threshold (e.g. 0.99). Thus the Main Net can be inspirited to pursue a higher fidelity. We find that the following reward function works well:

\begin{equation}\label{eq4}
	r=\begin{cases}
		100 \cdot F^{3}, & 0\leq F<0.99,  \\
		5000,            & 0.99\leq F\leq1.
	\end{cases}
\end{equation} 

For training the Main Net and evaluating the performance of our algorithm, we sample points on the Bloch sphere as the initial states. The training set contains
32 points that satisfy $\theta\in\{0,\pi/4,\pi/2,3\pi/4,\pi\}$ and $\varphi\in\{0,\pi/5,2\pi/5,3\pi/5,4\pi/5,\pi,$ $6\pi/5,7\pi/5,8\pi/5,9\pi/5\}$
on the Bloch sphere. The test set contains 320 points which is sampled
by inserting 2, 4 points into the intervals of the training set's
$\theta$ and $\varphi$, respectively $((5-1)\times2\times(10\times4)=320)$. Roughly speaking, these points are uniformly distributed on the surface of the Bloch sphere. It is worth stating that the hyperparameters used to train the Main Net here are slightly different for the preparation of the target states $|0\rangle$  and $|1\rangle$. The details of all hyperparameters for this algorithm has been listed in Table. ~\ref{tab1}.

\begin{figure}[ht]
	\subfigure[ ]{\includegraphics[width=7.5cm]{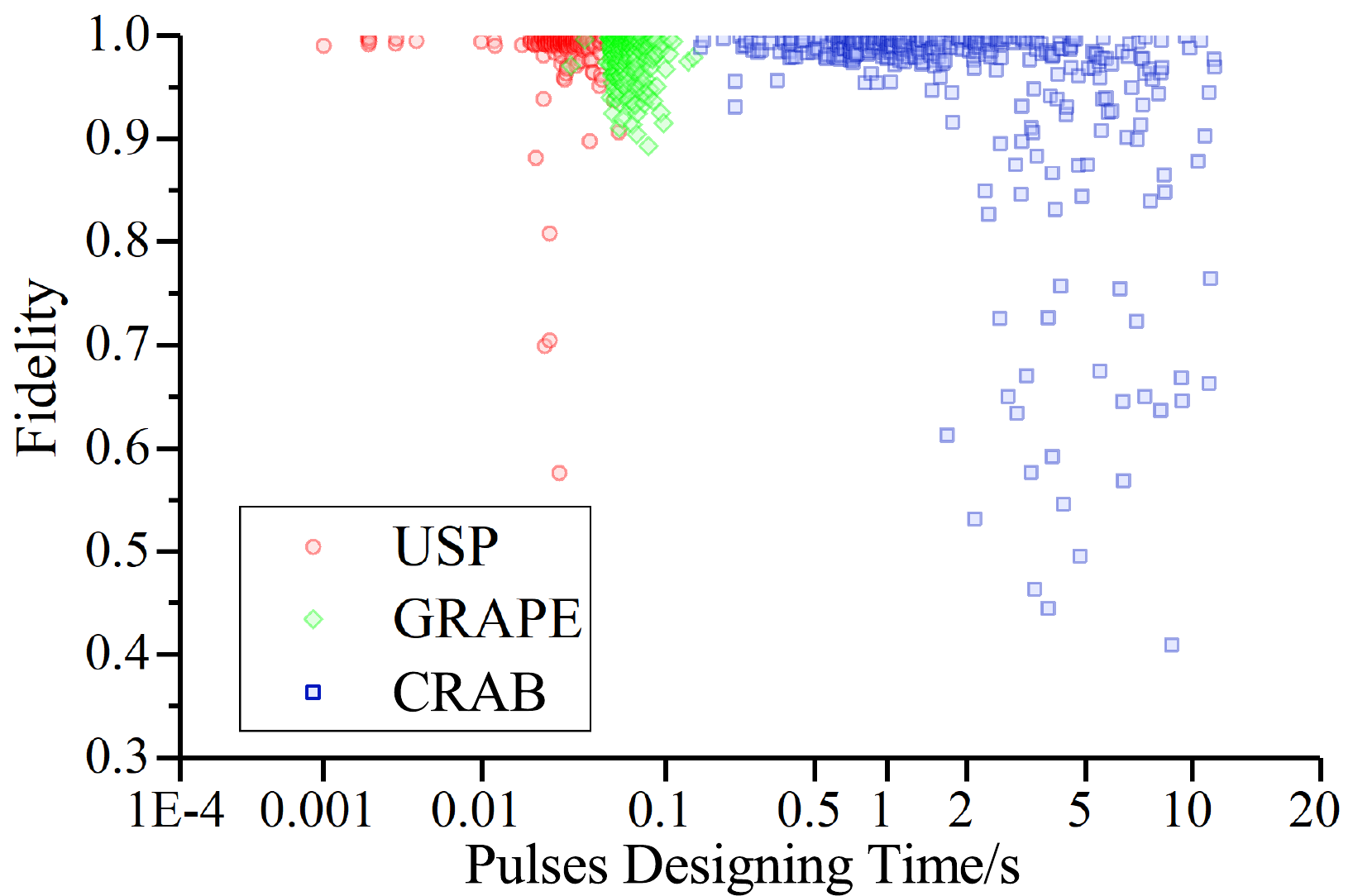}} 
	\subfigure[ ]{\includegraphics[width=7.5cm]{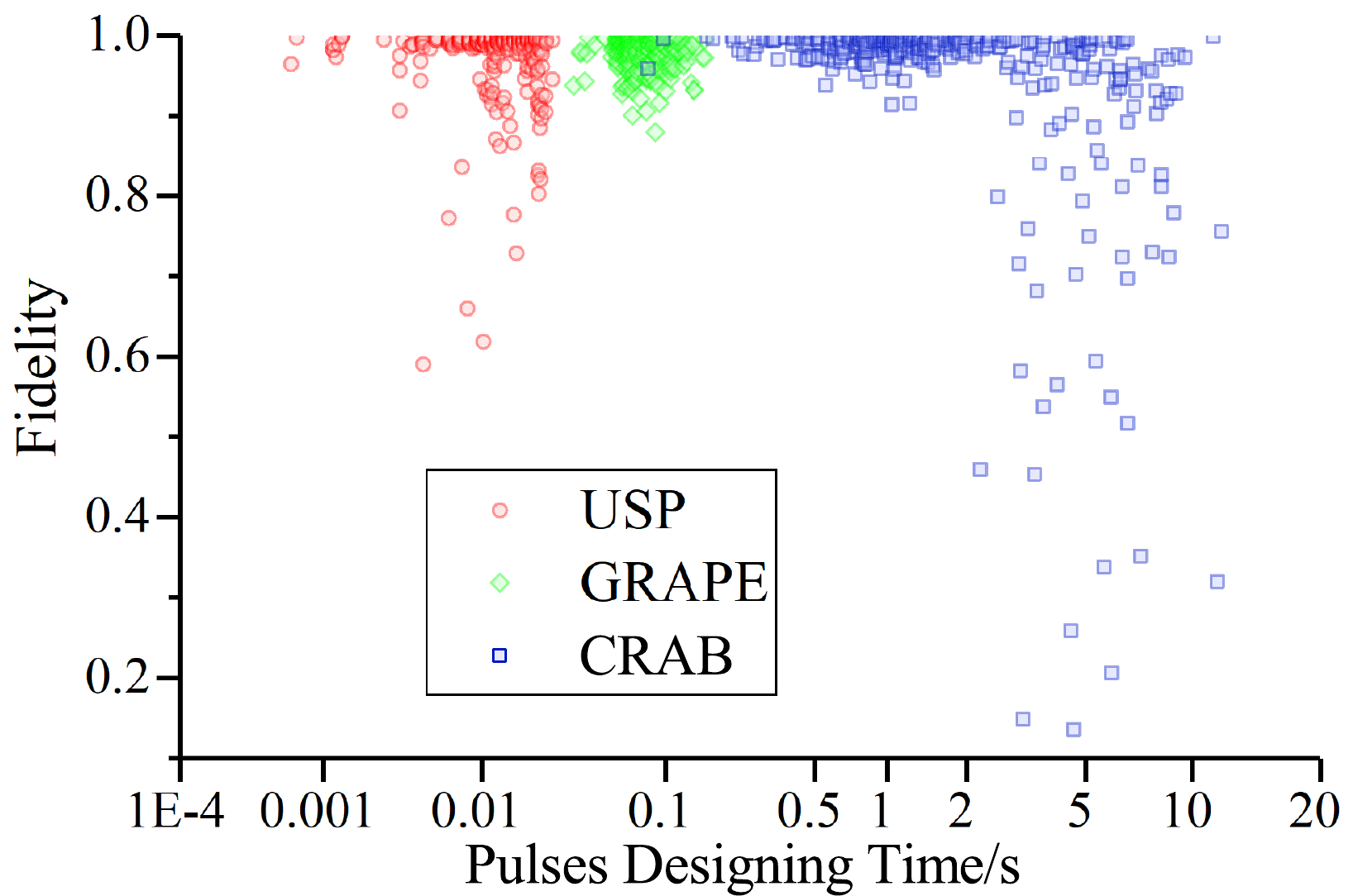}}
	\caption{The fidelity $F$ versus designing time distributions for preparing single-qubit state over 320 sampled tasks with different optimization algorithms. (a) Preparation of $|0\rangle$, where average fidelity $\overline{F}$ = 0.9875, 0.9812, 0.9468 and the average designing time $\overline{t}$ = 0.02950, 0.0651, 2.6457 with USP, GRAPE and CRAB. (b) Preparation of $|1\rangle$, where average fidelity $\overline{F}$ = 0.9740, 0.9708, 0.9388 and the average designing time $\overline{t}$ = 0.01167, 0.0789, 2.5285 with USP, GRAPE and CRAB.}\label{Fig.2}
\end{figure}

To compare the performance of our algorithm and the alternatives (GRAPE and CRAB), we plot their preparation fidelities of state $|0\rangle$ and $|1\rangle$ versus the corresponding runtime of designing the control trajectories in Figure~\ref{Fig.2} (a) and (b), respectively. 
To make a fair comparison, for the GRAPE and the CRAB, their continuous control strengths are discretized into the nearest allowed actions when the designing process is completed. The fidelities of the USP are the maximums that can be achieved within 40 time steps. Figure~\ref{Fig.2} shows that our USP algorithm outperforms the alternative optimization approaches both in terms of preparation quality and designing efficiency in discrete control space. Clearly, CRAB algorithm performs the worst, and GRAPE algorithm is in the middle. Note that the diversity of the runtime in designing control trajectories with USP stems from the difference of the required steps to finish these preparation tasks. A shorter designing time reflects a fewer steps, which corresponds to a faster control scheme in experiment.  While, the required steps of the traditional approaches are always fixed.

\begin{figure}[ht]
		\subfigure[ ]{\includegraphics[width=7.5cm]{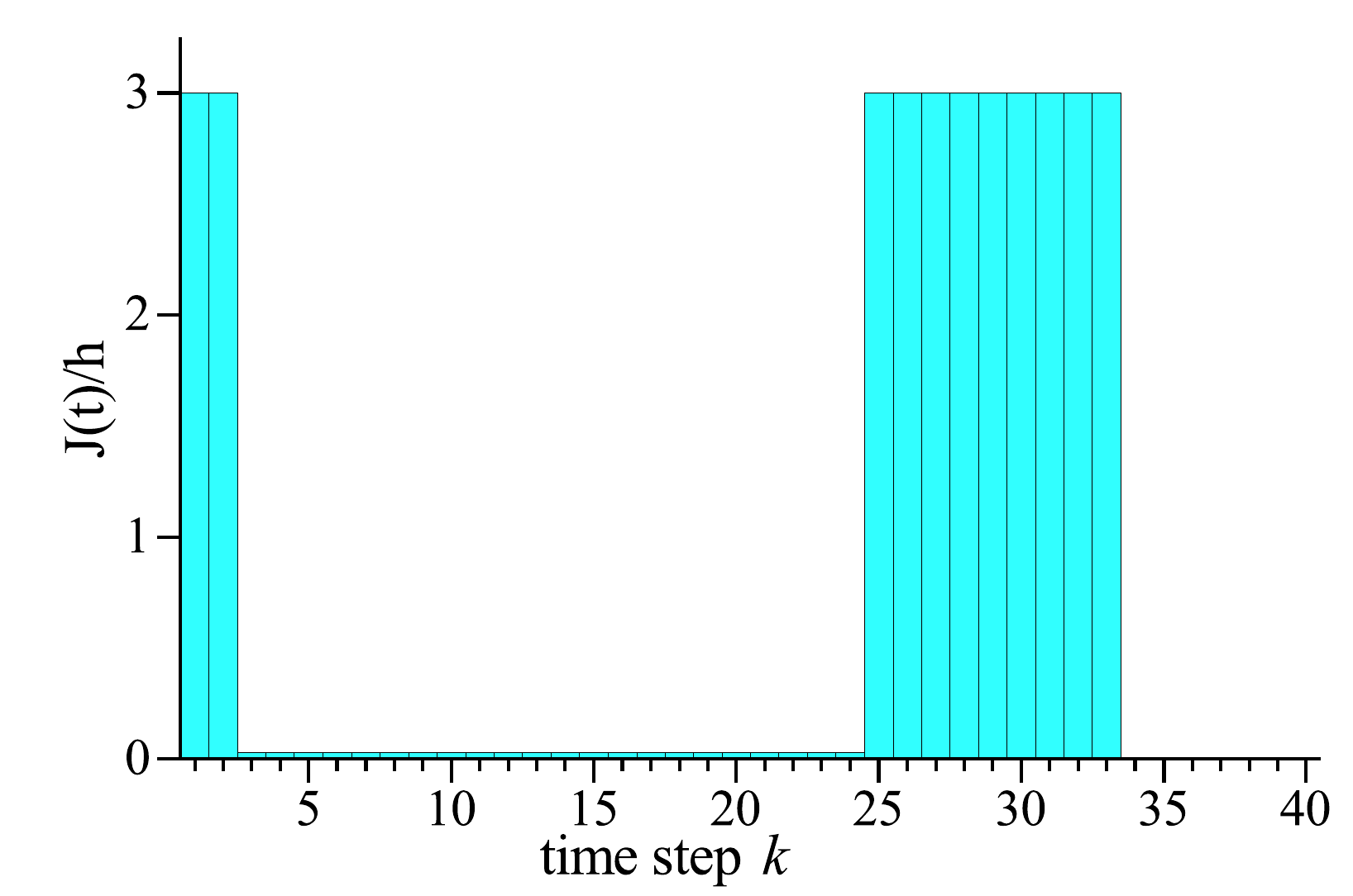}} 
	\subfigure[ ]{\includegraphics[width=7.5cm]{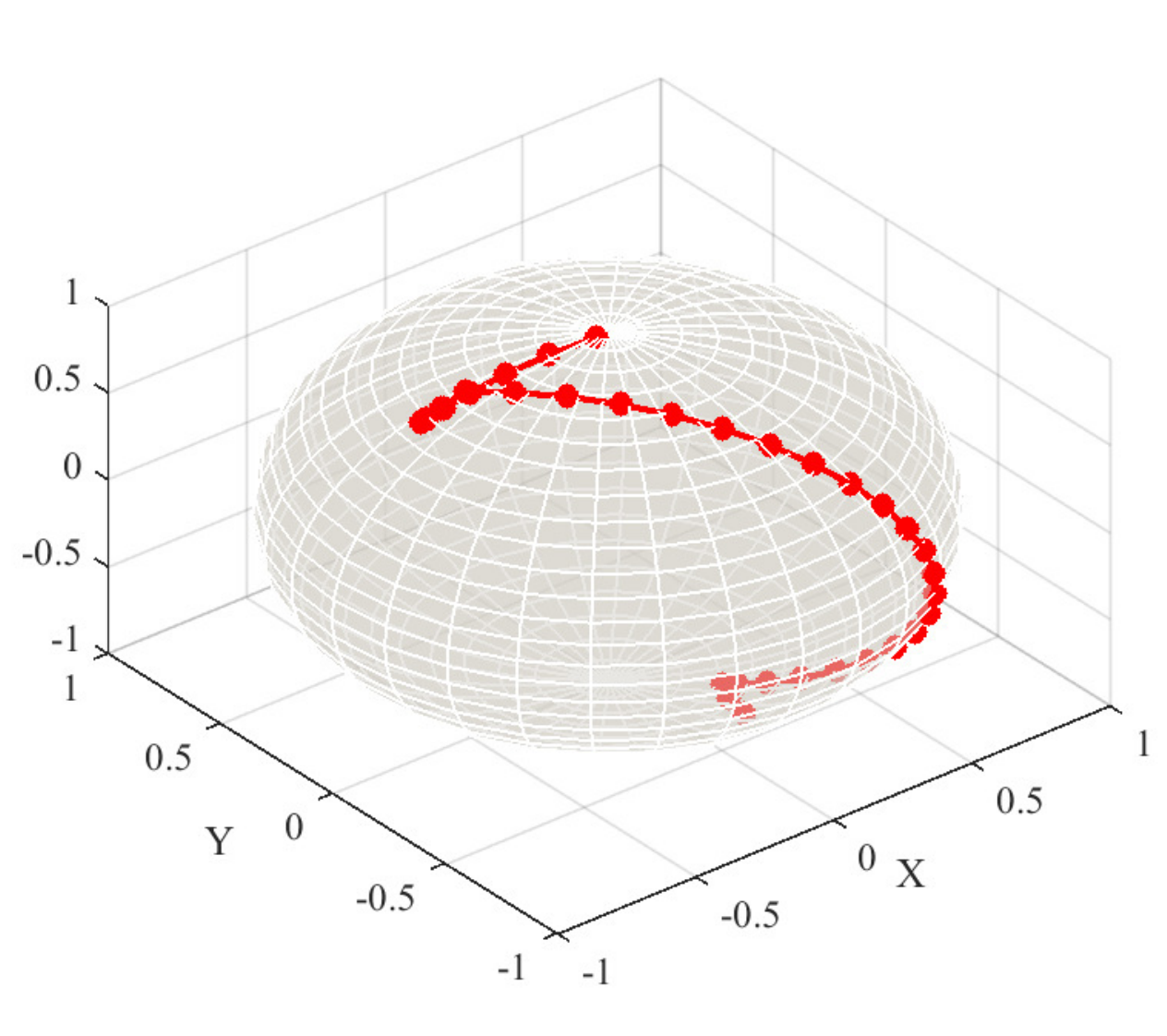}}
	\caption{(a) Control trajectory designed by the USP algorithm. The task is to reset the point $\theta=5\pi/6$, $\varphi=39\pi/25$ on the Bloch sphere to the target state $|0\rangle $. The pulses only take discrete values 0 and 3. This reset task is completed at time step 33. (b) The corresponding motion trail for the reset task on the Bloch sphere.}\label{Fig.3}
\end{figure}

To show the control trajectory designed by our USP algorithm visually, as an example we plot one in Figure~\ref{Fig.3} (a), where the position of the initial state on the Bloch sphere is $\theta=5\pi/6$, $\varphi=39\pi/25$ and the target state is $|0\rangle$.
It shows that the USP takes only 33 steps to complete the task.  The reason is that the DQN algorithm favors the policy with fewer time steps due to the discounted reward (See the details of the DQN described in Section~\ref{Sec.5}). 
In Figure~\ref{Fig.3} (b), we plot the corresponding
motion trail of the quantum state on the Bloch sphere during operations. It shows that the final quantum state reaches a
position that is very close to the target state $|0\rangle$ on the Bloch sphere. 

\subsection{Universal two coupled $S$-$T_0$ qubits state preparation}
\label{two-qubit state preparation}

Now we consider the preparation of the Bell state $(|00\rangle+|11\rangle)/\sqrt{2}$  \cite{nielsen2002quantum} from arbitrary initial states. The allowed pulse strength on each qubit is defined as $\{(J_1,J_2)|J_1,J_2 \in\{1, 2, 3, 4, 5 \}\}$.
The reward function is empirically set to be

\begin{equation}\label{eq5}
	r=\begin{cases}
		1000 \cdot F^{}, & 0\leq F<0.99,  \\
		5000,           & 0.99\leq F\leq1.
	\end{cases}
\end{equation}
The architecture of the Main Net employed in this task is slightly different from the one used for the manipulation of single-qubit and the detailed parameter values can be found in Table ~\ref{tab1}. The point set used to train and to test the algorithm contains 6912 points that are defined as $\{\left[a_1,a_2,a_3,a_4\right]^T\}$. $a_j=bc_j$ refers to the probability amplitude corresponding to the $j$th basis state. $b\in \{1,i,-1,-i\}$. $c_j$s together define points on a four-dimensional unit hypersphere, 

\begin{equation}
	\label{eq6}
	\begin{cases}
		\begin{array}{ll}
			c_{1}= & \mathrm{cos}\theta_{1},\\
			c_{2}= & \mathrm{sin}\theta_{1}\mathrm{cos}\theta_{2},\\
			c_{3}= & \mathrm{sin\theta_{1}\mathrm{sin}\theta_{2}\mathrm{cos}\theta_{3},}\\
			c_{4}= & \mathrm{sin}\theta_{1}\mathrm{sin}\theta_{2}\mathrm{sin}\theta_{3},
	\end{array}\end{cases}
\end{equation}
where $\theta_i\in\{\pi/8, \pi/4, 3\pi/8\}$. Note that the normalization condition is satisfied for each quantum state represented by these points.

\begin{figure}[ht] 
	\centering
	\includegraphics[width=8cm]{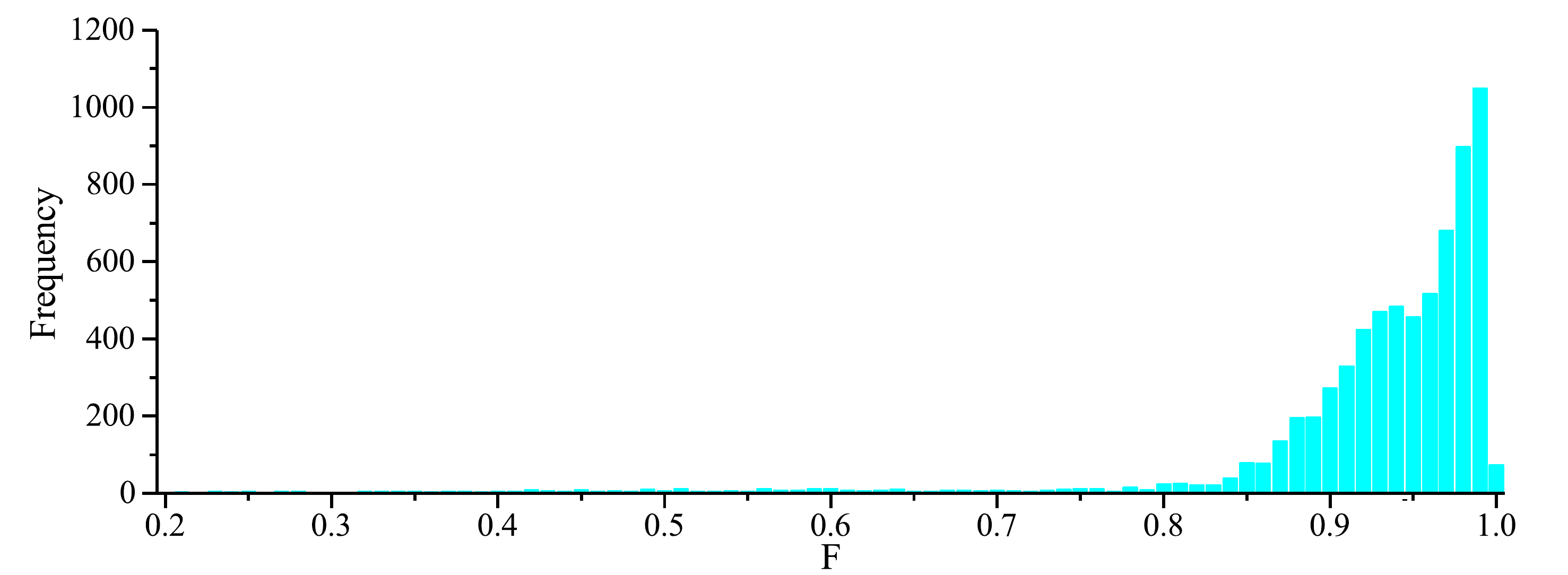}
	\caption{The frequency distribution of the fidelities of 6786 test points for the Bell state preparation.  The fidelities are the maximum values obtained during the maximum total operation time $T=10\pi$ 
		under control trajectories designed by our USP scheme.  The allowed actions satisfy $\{(J_1,J_2)|J_1,J_2 \in\{1, 2, 3, 4, 5 \}\}$ with the action duration $\mathrm{d}t=\pi/40$. The average fidelity is about 0.93.}
	\label{Fig.4}
\end{figure}

In the training process, we sample 126 points randomly from the point set as the training set. Each point in the training set is used to train the Main Net 100 episodes in turn. After training, the average fidelity of the Bell state preparation over all of the test points is about 0.93. The maximum total operation time is taken as $10\pi$ and it is discretized into 400 slices with pulse duration $\mathrm{d}t=\pi/40$.  In Figure~\ref{Fig.4}, we plot the frequency distribution of the fidelities of 6786 test points under control trajectories designed by our USP scheme in this two-qubit preparing task. It shows that although some fidelities are distributed unevenly between the interval $[0.2,0.9]$, the overall performance is good. In particular, these fidelities are concentrated at a high level ($F=0.99$).

\begin{table}[h!]
	\caption{\label{tab1} List of hyperparameters for USP.}
	\begin{tabular}{lccc}
		\\	\hline
		Parameters $\setminus$ Target state & $|0\rangle$ &  $|1\rangle$ &  Bell state \\\hline
		
		Allowed actions $J(t)$            &  0,1,2,3  & 0,1,2,3  & $^a$ \\
		number of points for training     &    32     &    32    &        126        \\
		number of points for testing      &    320    &   320    &       6786       \\
		Batch size  $N_{bs}$              &    32     &    32    &        320       \\
		Memory size $M$                   &   2000    &   3000   &      100000      \\
		Learning rate $\alpha$            &  0.0001   &  0.0001  &     0.000001     \\
		Replace period $C$                &    250    &   250    &        200       \\
		Maximum reward $r_{\mathrm{max}}$ &    5000   &   5000   &       5000       \\
		Reward discount factor $\gamma$   &    0.9    &   0.9    &        0.9       \\
		Number of hidden layers           &     2     &    2     &         3        \\
		Neurons per hidden layer          &   20/20   &   20/20  &    300/400/200   \\
		Activation function               &   Relu    &   Relu   &       Relu       \\
		$\epsilon$-greedy increment $\delta\epsilon$ &0.001&0.0001&     1/36000     \\
		Maximal $\epsilon$ in training $\epsilon_{max}$ & 0.99&0.99 &    0.99       \\
		Value of $\epsilon$ in testing    &     1    &     1     &         1        \\
		Maximum steps per episode         &    40    &     40    &       400        \\
		episodes per training point       &    100   &    100    &       100       \\
		Total time  $T$                      &  $\pi$   &   $\pi$   &      $10\pi$      \\
		Action duration $\mathrm{d}t$           & $\pi/40$ &  $\pi/40$ &      $\pi/40$    \\\hline
	\end{tabular}\\
	$^a$ The allowed actions of two-qubit operations satisfy $\{(J_1,J_2)|J_1,J_2\in \{1, 2, 3, 4, 5\} \}$.
\end{table}

\begin{algorithm}
	\caption{The pseudocode for training the USP algorithm.}\label{algo}
	\begin{algorithmic}
		\STATE Initialize the Experience Memory $D$ to empty.
		\STATE Randomly initialize the Main-network $\theta$.
		\STATE Initialize the Target-network $\theta^{-}$ by: $\theta^{-}\leftarrow\theta$.
		\FOR{point {\bf in}  training set} %
		\STATE Set the $\epsilon=0$.
		\STATE Set the initial state $s_{ini}$ according to the selected training point.
		
		\FOR{episode = 0, 100}
		\STATE Initialize the state $s=s_{ini}$.
		\WHILE{True}
		\STATE With probability $1-\epsilon$ select a random action $a_{i}$, otherwise $a_{i}=\mathrm{argmax}_{a}Q(s,a;\theta).$
		\STATE Set the $\epsilon=\epsilon+\delta\epsilon$, except $\epsilon=\epsilon_{max}$.
		\STATE Execute $a_{i}$ and observe the reward $r$, and the next state $s'$.
		\STATE Store experience $unit=(s,a_{i},r,s')$ in $D$.
		\STATE Select  batch size $N_{bs}$ of experiences units randomly from $D$.
		\STATE Update $\theta$ by minimizing the $Loss$ function.
		\STATE Every $C$ times of step, set $\theta^{-}\leftarrow\theta$.
		\STATE \textbf{break} if $r=r_{\mathrm{max}}$ or step$\geq T/\mathrm{d}t$.
		\ENDWHILE
		\ENDFOR
		\ENDFOR
	\end{algorithmic}
\end{algorithm}

\subsection{USP in noisy environments}
\label{noise analysis}

In the preceding section, we have studied the USP without considering the surrounding environment. However the qubits will suffer from a variety of fluctuations in a practical experiment, such as the magnetic noise \cite{malinowski2017notch}. It comes from the uncontrolled
\begin{figure}[H]
	\centering	
	\subfigure[ ]{\includegraphics[width=7cm]{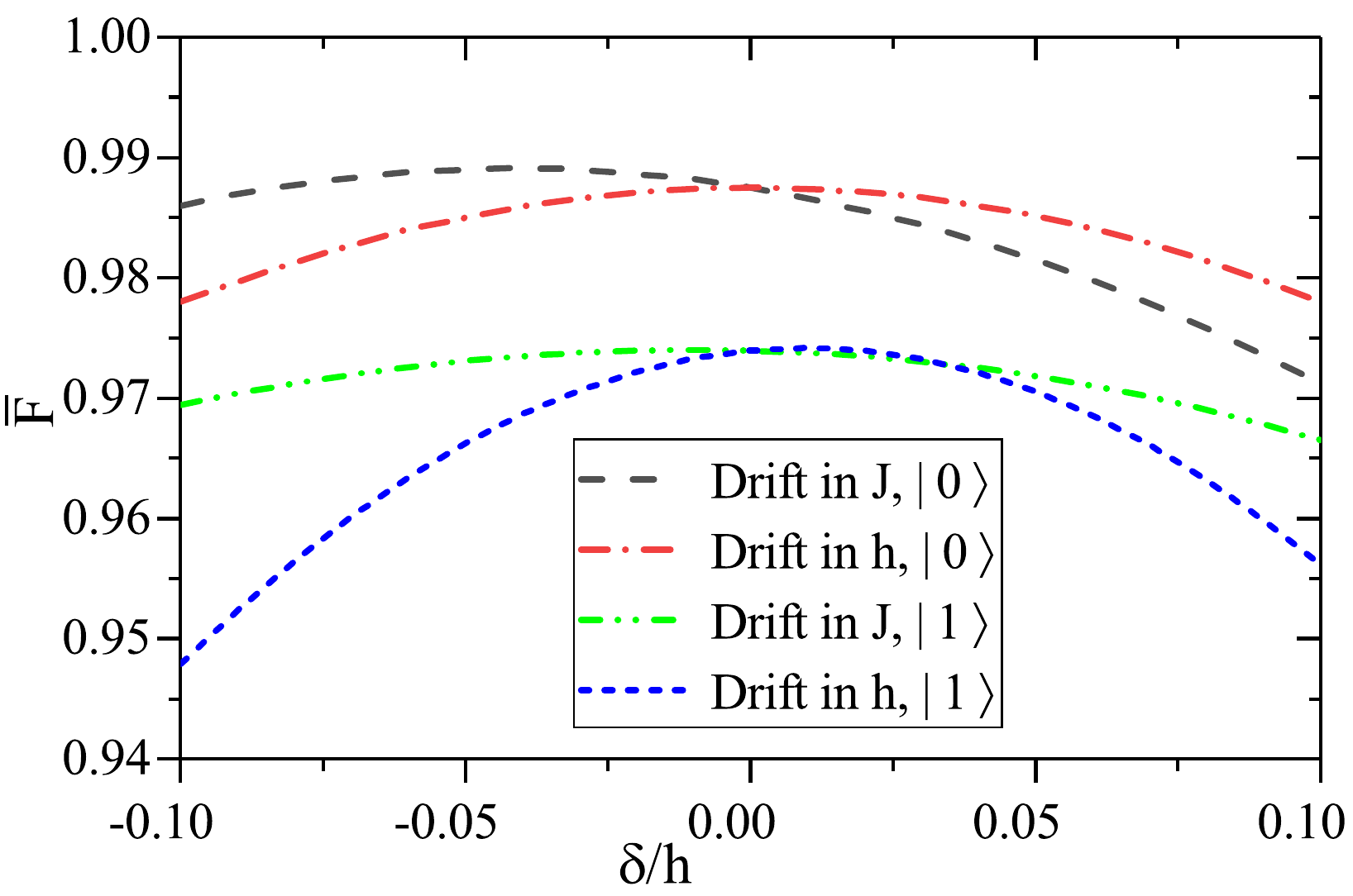}} 
	\subfigure[ ]{\includegraphics[width=7cm]{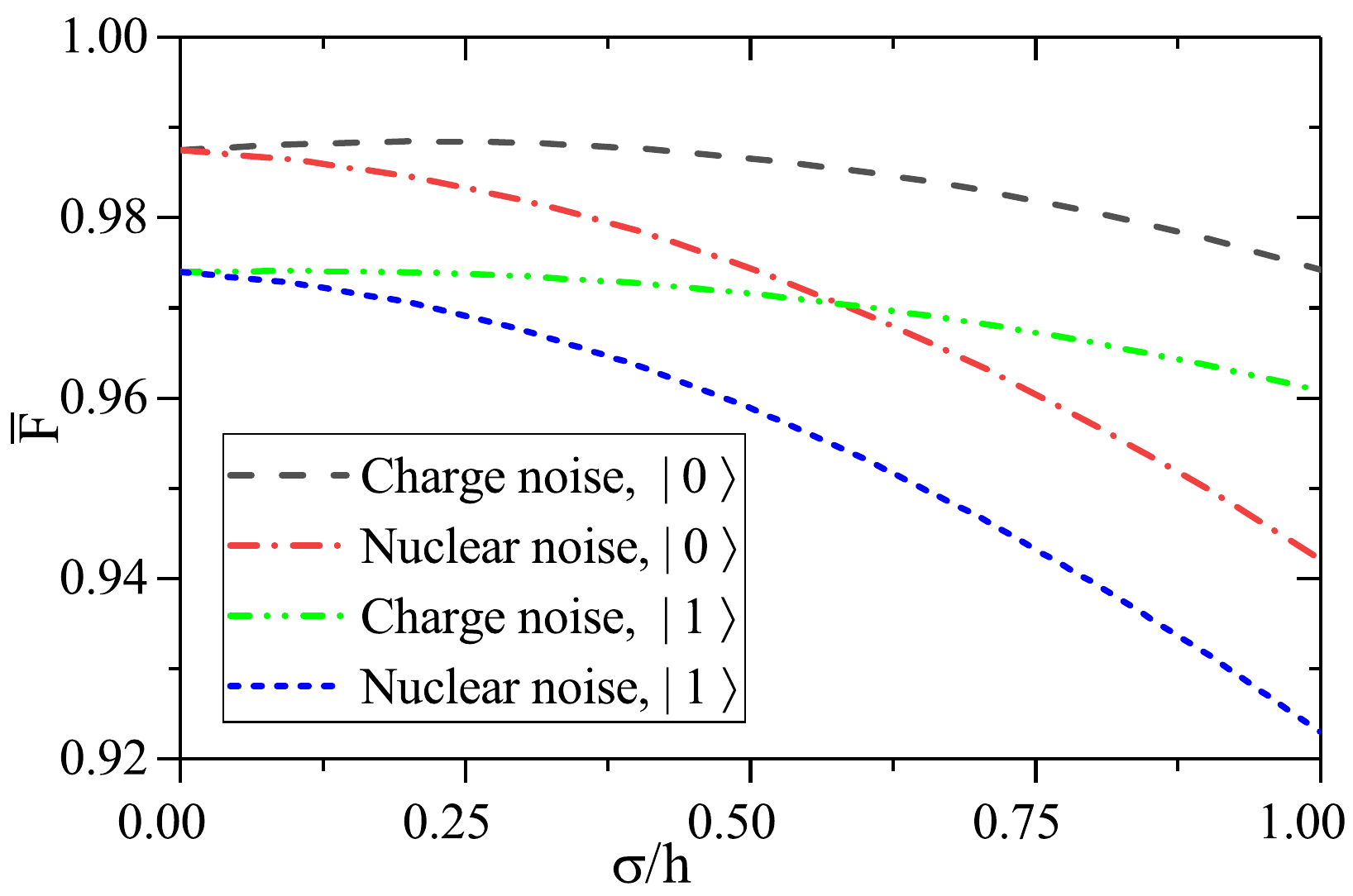}}	
	\subfigure[ ]{\includegraphics[width=7cm]{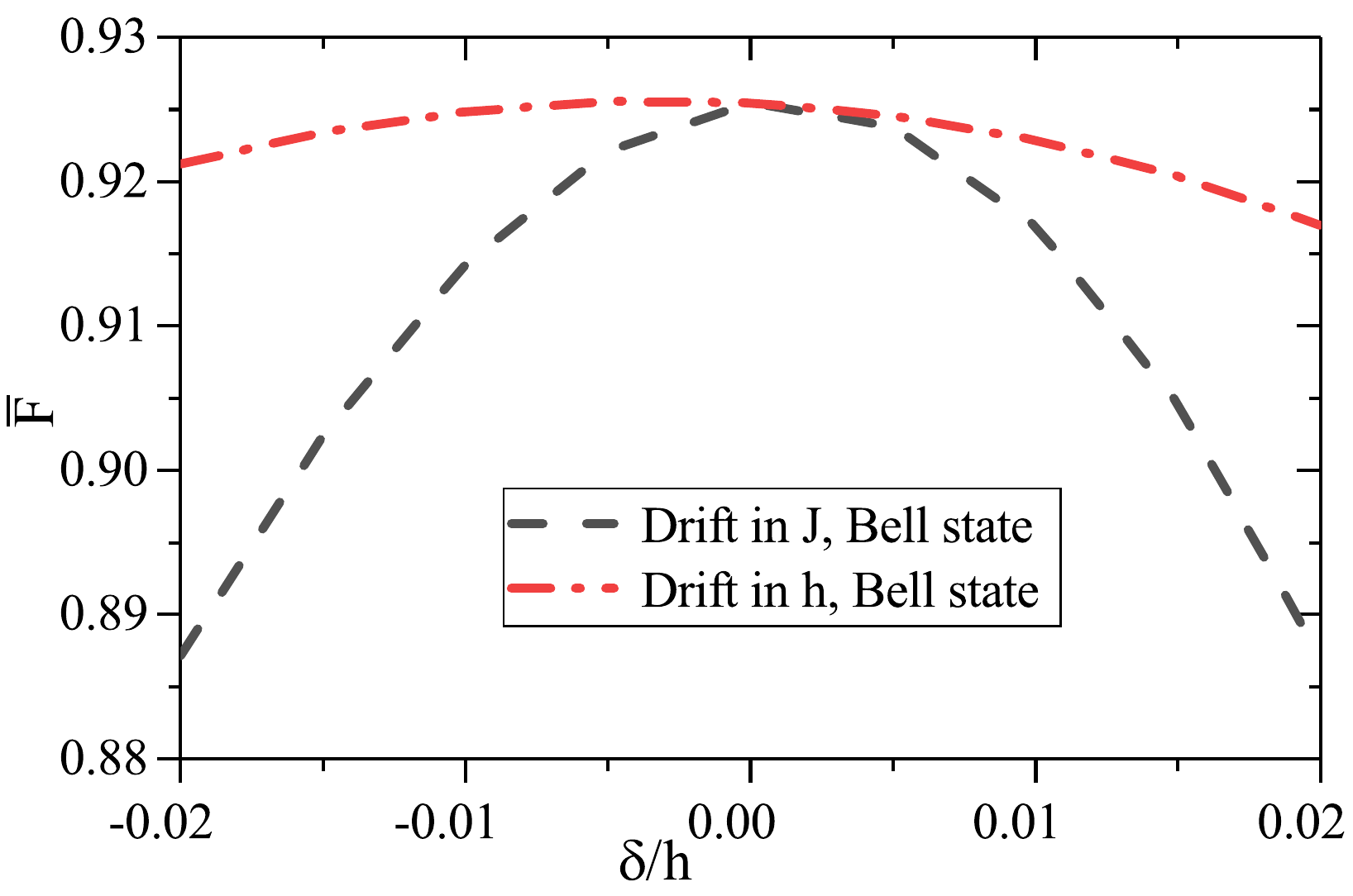}} 
	\subfigure[ ]{\includegraphics[width=7cm]{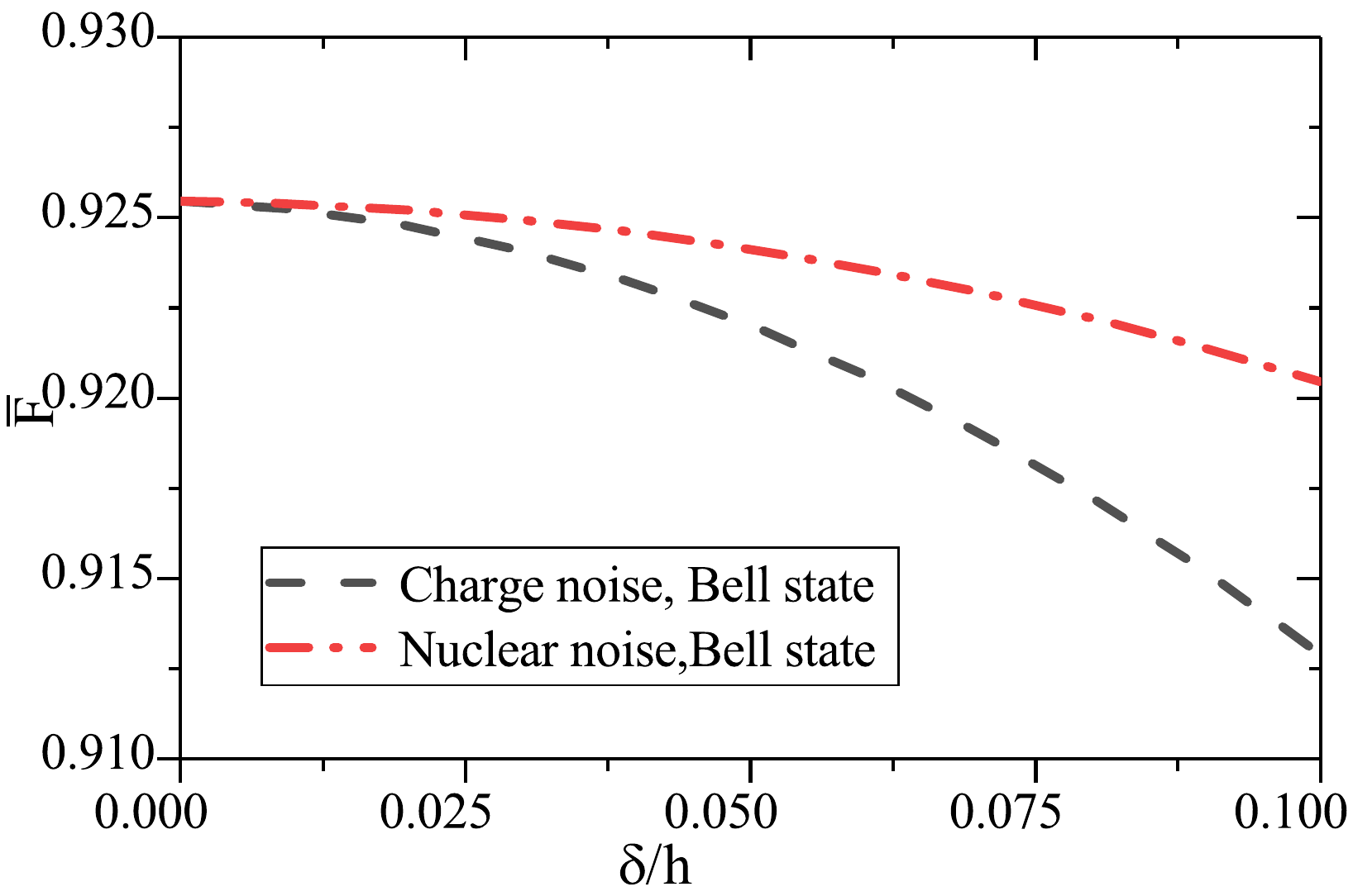}}	
	\caption{Average fidelity of the single- and two-qubit state preparation with USP algorithm over all test points vs amplitudes of various imperfections: (a) the static drifts in $J$ and $h$ for single-qubit target state $|0\rangle$ and $|1\rangle$; (b) the dynamic charge noise and nuclear noises for single-qubit target state $|0\rangle$ and $|1\rangle$; (c) the static drifts in $J$ and $h$ for the Bell state; (d) the dynamic charge and nuclear noises for the Bell state.}\label{Fig.5}
\end{figure}
hyperfine coupling with spinful nuclei of the host material.
Next we will study the control trajectories designed by our USP algorithm under two types of static drifts and two types of dynamic fluctuations. 
Here we point out that all of these imperfections are integrated into the system's evolution after the control trajectories have been designed by our Main Net, which is trained on a clean model. This is reasonable since normally the environment is unpredictable.  

The static drifts could be caused by the imperfections of the external fields. For the control of single-qubit, these drifts can be written as an additional term $\delta \sigma_{z}$ or $\delta \sigma_{x}$ in the Hamiltonian (\ref{eq1}),  where $\delta$ is the amplitude of the drifts. For the manipulation of two-qubit, static drifts can be taken by replacing the term $J_i$ (or $h_i$) with $J_i+\delta_i$ (or $h_i+\delta_i$), where $i\in\{1,2\}$ in the Hamiltonian (\ref{eq3}).
The dynamic fluctuations include charge and nuclear noises  \cite{petta2005coherent,barnes2012nonperturbative,nguyen2011impurity}. They originate from the environment and can be represented by replacing the term $J(t)$ (or $h$) with $J(t)+\delta(t)$ (or $h+\delta(t)$) in the Hamiltonian (\ref{eq1}) for single-qubit, or by replacing  $J_i$ (or $h_i$) with $J_i+\delta_i(t)$ (or $h_i+\delta_i(t)$) in the Hamiltonian (\ref{eq3}) for two-qubit. Here $\delta(t)$ and $\delta_i(t)$ are sampled randomly from a normal distribution $N(0,\sigma^{2})$. In the simulation, the time-dependent fluctuations $\delta(t)$ and $\delta_i(t)$ are taken as a constant within each time step. The magnitude of $\delta(t)$ or $\delta_i(t)$  grows with an increasing parameter $\sigma$ of the normal distribution function. 

For single-qubit state preparation, the average fidelity $\overline{F}$ as a function of the fluctuation magnitude in the presence of four types of fluctuations are plotted in Figure~\ref{Fig.5} (a) and (b). Here the average fidelity $\overline{F}$ is calculated over all test points. Figure~\ref{Fig.5} (a) and (b) show that the average fidelity does not change significantly with considered imperfections. From Figure~\ref{Fig.5}, we find that in the analyzed parameter window $\overline{F}$ with target state $|0\rangle$ is always higher than with target state $|1\rangle$. Figure~\ref{Fig.5} also shows that $\overline{F}$ is most severely affected by the drift in $h$ (or nuclear noise) with target state $|1\rangle$. For two-qubit Bell state preparation, we assume that the amplitudes of the static noises on the two qubits are identical and the dynamic noises are different. Now $\delta_1 = \delta_2$ and  $\delta_1(t) \neq \delta_2(t)$. For the two qubits, the dynamic noises are sampled from a same dynamic normal distribution $N(0,\sigma^{2})$ individually. Figure~\ref{Fig.5} (c) and (d) plot the average fidelity of the Bell state preparation over 6786 test points in the presence of noises. They show a similar behavior as the single-qubit preparation case. The average fidelity also does not change significantly with considered imperfections. The difference is that $\overline{F}$ under the charge noise is smaller than the nuclear noise case. 
Note that the best fidelities can even be obtained in some non-zero imperfections (static drift in $J$) from Figure~\ref{Fig.5} (a) and (c). That is to say, certain noises can be helpful to boost the fidelity due to subtle adjustments on $J$. A possible explanation may be the
limitation of the $J$ value in our calculation. We believe that there is still a room for the achievement of better performance by employing more allowed actions $J$ and more deliberate Zeeman energy spacing $h$, just as what these noises do. Of course, more sufficient training on Main Net is also helpful for the enhancement of the fidelity.

Given the limitations of quantum computing hardwares presently accessible,	we simulate quantum computing on a classical computer and generate	data to train the network. Our algorithms are implemented with PYTHON 3.7.9, TensorFlow 2.1.0 and QuTip 4.5 and have been run on an four-core 1.80 GHz CPU with 8 GB memory. Details of the running environment of the algorithm can be found in the Section \textbf{Availability of data and materials}.
The runtime for the training process of algorithms are about a few minutes in the single-qubit case and several hours in the two-qubit case.

\section{Conclusions}
\label{Sec.4}
Precise and efficient quantum state preparation is crucial for quantum information processing. In this paper, we proposed an efficient scheme to generate appropriate control trajectories to  reset arbitrary single- or two-qubit states to a certain target state with the aid of deep RL. 
Our scheme has the advantage that once the network is well trained, it works for arbitrary initial states and does not require training again. 
Taking the control of spin $S$-$T_0$ qubits in semiconductor DQDs as an example, the evaluation results show that our scheme outperforms traditional optimization approaches with both preparation quality and pulses designing efficiency.
Moreover, we find that the control trajectories designed by our scheme exhibit strong robustness against various types of static and dynamic noises. Although we only consider the single and two-qubit state preparation in semiconductor DQDs, this scheme can be extended to a wide variety of quantum control problems.

\section{Deep reinforcement learning and deep Q network}
\label{Sec.5}
In this section, we will introduce the deep RL and DQN algorithm, which underlie our USP scheme.

\begin{figure}[H] 
	\centering
	\includegraphics[width=7cm]{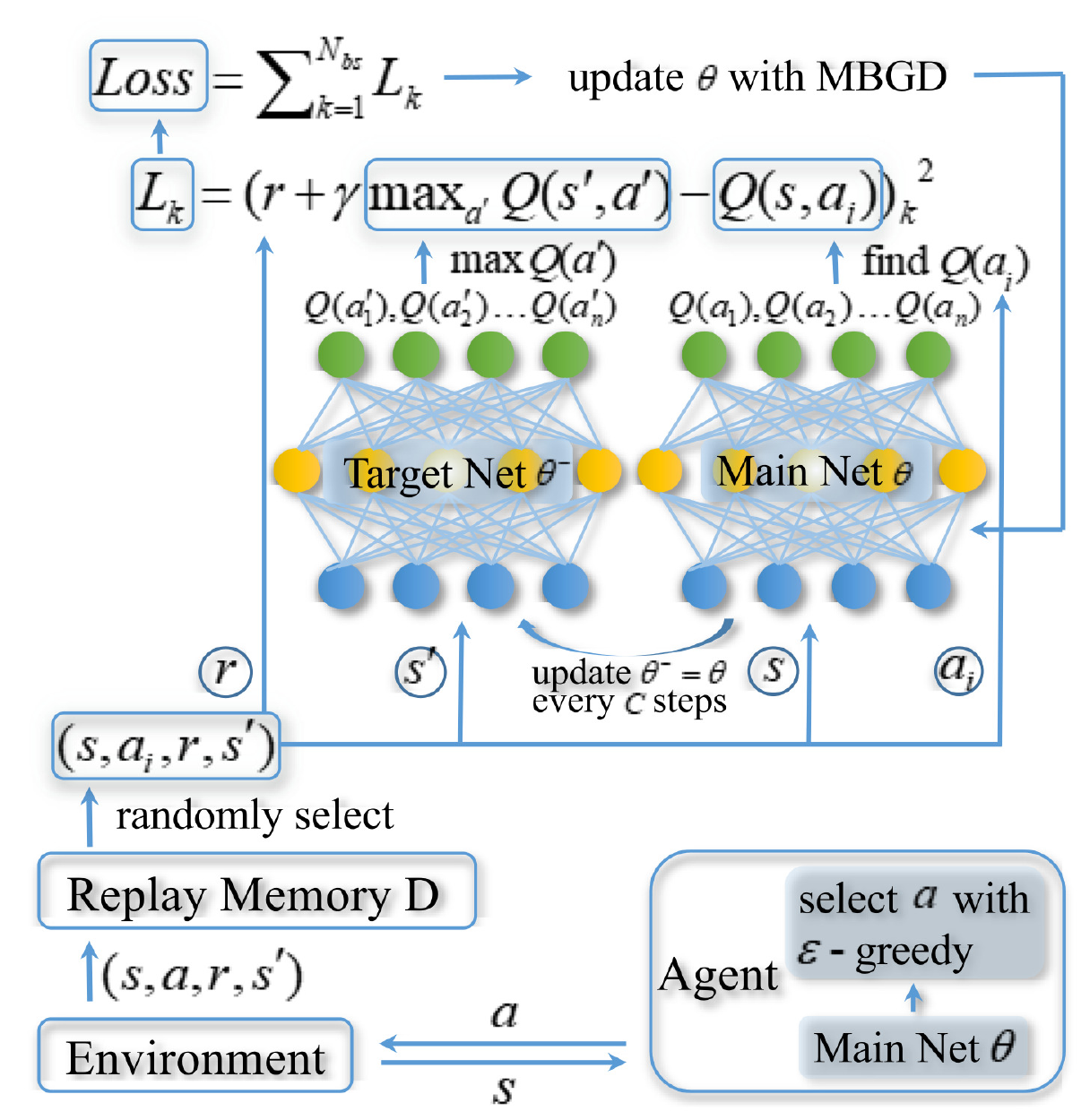}
	\caption{Schematic for the DQN algorithm. See the main text of the Section~\ref{Sec.5} for the details of the algorithm.}
	\label{Fig.6}
\end{figure}

The deep RL combines the deep learning algorithm that is good at nonlinear fitting and the RL algorithm that is expert in dynamic programming problems \cite{shalev2014understanding,goodfellow2016deep,sutton2018reinforcement}.

In RL, an Agent is generally used to represent an object with decision-making and action capability, such as a robot. We consider a Markov decision process in which the next state depends only on the current state as well as the action performed by the Agent and has no relation with the past states \cite{sutton2018reinforcement}. In the interaction between the Agent and the Environment, the current state $s$ of the Environment will be changed to another next state $s'$, after the Agent selecting and performing an action $a_{i}$ chose from the set of allowed actions $a=\{a_{1},a_{2}\cdots,a_{n}\}$ at  time  $t$. In return, the Environment also gives a feedback, or immediate reward $r$ to the Agent. A Policy $\pi$ represents which action the Agent will be chose in a given state, i.e., $a_{i}=\pi(s)$. The process is defined as an episode in which the Agent starts from an initial state until it completes the task or terminates in halfway.

The total discounted reward $R$ gained in an $N$-steps episode can be written as \cite{sutton2018reinforcement}:

\begin{equation}\label{eq7}
	R\!=\!r_{1}\!+\!\gamma r_{2}\!+\!\gamma^{2}r_{3}\cdots\!+\!\gamma^{N-1}r_{N}\!=\!\sum_{t=1}^{N}\gamma^{t-1}r_{t},
\end{equation}
where $\gamma$ is a discount factor within the interval $[0,1]$, which indicates that the immediate reward $r$ discounts with the steps increasing. Assume the Agent will get a big  reward $r_{\mathrm{max}}$ when it reaches the target state and then ends the current episode. Because the discounted $r$, the Agent tends naturally to get that final bonus by completing the task as quickly as possible.	
The goal of the Agent is to maximize $R$, because a greater $R$ implies a better performance of the Agent. To determine which action should to be chose in a given state, we introduce the action-value function, which is also named $Q$-value \cite{watkins1992q}:

\begin{equation}\label{eq8}
	Q^{\pi}(s,\!a_{i})\!=\!E\left[r_{t}\!+\!\gamma r_{t+1}\!+\!\cdots|s,\!a_{i}\right]\!=\!E\left[r_{t}\!+\!\gamma Q^{\pi}(s'\!,\!a')|s,\!a_{i}\right].
\end{equation}
The $Q$-value indicates the expectation of $R$, which the Agent will get after it executing an action $a_{i}$ in a given state $s$ under the policy $\pi$, and this value can be obtained iteratively according to the $Q$-values of the next state. Because there are multiple allowed actions can be chosen in each state, and different actions will lead to different next states, it is a time-consuming task to calculate $Q$-values in a multi-step process. To reduce the overhead, there are various algorithms used to calculate approximations of that expectation, such as $Q$-learning \cite{watkins1992q} and SARSA \cite{sutton2018reinforcement}.

In $Q$-learning, the current $Q(s,a_{i})$ value is obtained by the $Q$-value of the next state's ``best action'' \cite{watkins1992q}:

\begin{equation}\label{eq9}
	Q(s,\!a_{i})\!\leftarrow\! Q(s,\!a_{i})\!+\!\alpha\!\left[r_{t}\!+\!\gamma\mathrm{max}_{a'}Q(s'\!,\!a')-Q(s,\!a_{i})\right],
\end{equation}
where $\alpha$ is the learning rate, which affects the convergence of this function. The part of $Q_{target}(s',a')=r_{t}+\gamma\mathrm{max}_{a'}Q(s',a')$ is called the $Q_{target}$ value. All the $Q$-values of different states and actions can be recorded in a so-called $Q$-Table. With a precise $Q$-Table, it is easily to identify which action should be chose in a given state.
However,  on the one hand, we need the best action to calculate iteratively the $Q$-value; on the other hand, we must know all the $Q$-values to determine which action is the best. To solve this dilemma of ``exploitation'' and ``exploration'', we adopt the $\epsilon$-greedy strategy in choosing action to execute, i.e., choose the action corresponding to the current maximum $Q$-value with a probability of $\epsilon$ to calculate $Q$-value efficiently, or choose an action randomly with a probability of $1-\epsilon$ to expand the range of consideration. At the beginning, since it is not known that which action is the best one in a certain state, the $\epsilon$ is set to be $0$ to explore as many states and actions as possible. When sufficient states and actions are explored, that parameter gradually increases with the amplitude of $\delta\epsilon$ until to $\epsilon_{max}$, which is slightly smaller than $1$, to calculate the $Q$-values efficiently.

For an Environment with a large number or even an infinite number of states, the $Q$-Table would be prohibitively large. To solve this ``dimensional disaster'', we can substitute this table with a multi-layer neural network. After learning, the network will be capable to match  a suited $Q$-value to each action after be fed with a certain state. The deep $Q$ network algorithm  (DQN) \cite{mnih2013playing,mnih2015human} are based on the Equation~(\ref{eq9}). A network, the Main Net $\theta$, is used to predict the term $Q(s,a_i)$, and another network, the Target Net $\theta^-$ is used to predict the term $\mathrm{max}_{a'}Q(s',a')$ in Equation~(\ref{eq9})  respectively. In order to ensure the ability of generalization, the data used to train the Main Net must meet the assumption of independent and identically distributed, i.e. each sample of the dataset is independent of another and the training and test set are identically distributed. So we adopt the experience memory replay strategy \cite{mnih2015human}: the Agent could get an experience unit $(s,a,r,s')$ at each step. After numerous  steps, the Agent will collect a lot of such units that can be stored in an Experience Memory $D$ with capacity of Memory size $M$. In the process of training, the Agent randomly samples batch size $N_{bs}$ of experience units from the Experience Memory to train the Main Net at each time step.

Notice that to ensure the stability of the algorithm only the Main Net is trained in every time step by minimizing the $Loss$ function:

\begin{equation}\label{eq10}
	Loss=\frac{1}{N_{bs}}\sum_{i=1}^{N_{bs}}\left([r+\gamma\mathrm{max}_{a'}Q(s',a')]_{ i}-Q\left(s,a\right)_{i}\right)^{2},
\end{equation}
where $N_{bs}$ is the sample batch size through mini-batch gradient descent (MBGD) algorithm \cite{goodfellow2016deep,mnih2013playing,mnih2015human}. While the Target Net $\theta^{-}$ is not updated	in real time, instead, it copies the parameters from the Main Net $\theta$ every $C$ steps.
A schematic of this DQN algorithm is shown in Figure~\ref{Fig.6}.

\section*{Acknowledgements}
We would like to thank Xin-Hong Han, Jing-Hao Sun and Chen Chen for useful discussions.

\section*{Funding}%
This work was supported by the Natural Science Foundation of China(Grant Nos. 11475160, 61575180), and the Natural Science Foundation of Shandong Province (Grant No. ZR2014AM023).

\section*{Availability of data and materials}
The code, running environment of algorithm and all data supporting the conclusions of this article are available from the corresponding author on reasonable request or on  Github repository under MIT License (\url{https://github.com/Waikikilick?tab=repositories}).

\section*{References}
\bibliography{References}
\end{document}